# Full Transparency in DBI frameworks

Vlad Constantin Crăciun[1]
Andrei-Cătălin Mogage[1]
Dorel Lucanu[1]
[1] Alexandru Ioan Cuza" University of IAȘI, Department of Computer Science
{vex.christmas, andrei.mogage, dorel.lucanu}@gmail.com

***Abstract*—Following the increasing trends of malicious applications or cyber threats in general, program analysis has become an ubiquitous technique in extracting relevant features. The current state of the art solutions seem to fall behind new techniques. For instance, dynamic binary instrumentation (DBI) provides some promising results, but falls short when it comes to ease-of-use and overcoming analysis evasion. In this regard, we propose a two-fold contribution. First, we introduce COBAI (Complex Orchestrator for Binary Analysis and Instrumentation), a DBI framework designed for malware analysis, prioritizing ease-of-use and analysis transparency, without imposing a significant overhead. Second, we introduce an aggregated test suite intended to stand as a benchmark in determining the quality of an analysis solution regarding the protection against evasion mechanisms. The efficiency of our solution is validated by a careful evaluation taking into consideration other DBI frameworks, analysis environments and the proposed benchmark.**

## I. Introduction

The current research proposes a framework for binary analysis, having the main focus set on malicious applications that are evasive, in attempt to overcome their capabilities of escaping the analysis environment or the refusal of execution under such scenarios.

Binary analysis is a tedious domain, which demands a continuous process [56], [52] of designing and developing new protocols, mechanisms and techniques, in order to survive and overcome industry changes. Apart from that, malicious applications raise the bar even further, implementing various techniques to slow or deter the analysis [57], [32].

The main challenge raised by the analysis of evasive malware is that the analysis environment must be indistinguishable from an ordinary one, e.g. belonging to a regular user and which lacks any analysis tools. Therefore, the analysis solution should be able to provide a high level of *transparency*. By analysis environment we refer to the entire scope where the application is analyzed, comprising the operating system, virtualization / emulation technologies, debuggers, additional static or dynamic analysis tools and so on. However, in order to be able to analyze evasive malware, one may need to adjust its environment so that any analysis artifacts are concealed. The flow of installation, preparation and configuration leads to a secondary problem: ease of use.

The increasing trend on techniques and strategies for analysis evasion [27] have consisted the main motivation for the development of a new framework for binary analysis instead of relying on already existing ones. As the evaluation in Section III reflects, the state of the art candidates are not entirely suitable, without a significant effort, for a proper analysis of highly evasive malware.

### A. Dealing with DBI Transparency

A dynamic binary instrumentation framework is a popular solution for binary analysis where the code can be arbitrarily executed. As a result, a high level of control may be obtained and this allows a better manipulation of the application according to the purpose of the analysis. As a consequence, unfortunately, attackers started developing various strategies to detect and counter-attack DBI solutions. There is a general lack of countermeasures against transparency detection and attacks [35]. Moreover, it is of utmost importance to be able to verify the transparency properties of an analysis solution or to compare multiple tools in this regard, therefore an aggregated set of transparency tests is an essential factor.

A generic DBI framework, for instance, is usually correct while instrumenting benign code, but it does not totally prevent the detection of the analysis process while instrumenting malicious code. It does not deploy mechanisms to trick the application into considering it is not detected, which is not a direct feature of the instrumentation. We consider that the transparency aspect also relates to the ease of use earlier discussed, since the instrumentation is correct, but the desired results are different. This leads to either integrating additional projects purposely created to aid in this type of tasks, or having to develop new ones.

For instance, the users, whose purpose is to analyse an evasive malware with an existing DBI solution, might be enforced to resort to at least one of following two steps: integrating an existing tool / plugin whose purpose is to make the engine transparent or to develop it by themselves. Both solutions are highly sensitive for engines without a special care for transparency, as they might prove to be difficult to integrate, break backwards compatibility, or even generate side effects, including crashes. Alternatively, users may increase the DBIs transparency leveraged by specific instrumentation hooks. These hooks allows slightly shifts of the application behavior, such that some of the DBI exposed resources may become invisible. This type of mitigation include instruction-level or API-level access to incomplete virtualized DBI resources. However, there are cases where only DBI developers may assist with such functionalities, as they either were not present in the first place, or they were not sufficiently tested.

When the application being instrumented implements analysis evasion techniques, it either targets some resources that are not correctly virtualized, or it pushes the limits of available resources and performance counters, forcing the application to either generate execution crashes (as a side effect of exceeding some physical boundaries) or increase some overheads in a visible way. The subject of DBI virtualization was not enough debated along the road. A question to answer is whether the full virtualization should be handled by the DBI engine, or should users be responsible for it. To our surprise, seeing the large number of DBI tools designed to increase the DBI transparency, we concluded that indirectly the DBI developers made users responsible for the missing virtualization features. Our statement is confirmed by a considerable community effort, where projects like SoK [29], [15], PinVMShield [13], BluePill [4], JuanLesPIN [22], [46], all attempt to patch the virtualization gap in PIN [45] DBI framework. This lack of virtualization did not raised any concerns when the DBI frameworks were first developed about two decades ago, mainly because the benign application used to test the DBIs required actually a lower amount of resource-virtualization, compared to what a DBI was capable to provide. In the meantime, hackers increased the awareness of their malicious applications, implementing countless analysis-evasion scenarios, meant to disrupt any attempt to reverse-engineer the binaries. Their efforts contributed with interesting testing scenarios questioning the virtualization responsibilities and also raised some architectural issues such that developers behind the DBI frameworks may have difficulties handling them correctly. We believe that apparently making users responsible for the virtualization gap, is just an unconventional solution to provide some architectural fixes without forcing developers to deal with the whole process of integrating those fixes at the engine level. Malicious applications usually exploit either the boundaries of the virtualization rate (known uncovered virtualization aspects, or virtualization issues unknown to the public) or the increased resources overhead (time, CPU, memory) added by the presence of a DBI.

An analyst may use sandbox technologies specially designed for malware analysis, such as Cuckoo sandbox [5], Any Run [2], or Hybrid Analysis [9]. Unlucky, the attackers adapted on this field as well by: refusing to execute under a VM or a sandbox, trying to escape the debugger, attempting to kill the processes of analysis tools, etc. This challenge is of the utmost importance, as a detected analysis environment will not be able to reveal the necessary level of details. For instance, regarding the countermeasures for OS environment transparency that Cuckoo sandbox should handle, [47] (pg. 113) states: *"It is difficult to make all the sensors get false values, but at least we can try to reduce the detection. It is said that we can reduce about 90 percent of it."*. The authors statement is relative to the API hooking analysis technology present in Cuckoo Sandbox monitoring module. Projects like BluePill [4], PinVMShield [13] and JuanLesPIN [46], [22] attempt to extend the DBI virtualization functionalities above the DBI, including OS resources. If the previously discussed monitoring module relies only on API hooking to reduce about 90% of a sandbox fingerprints, a DBI is more suitable for this task as its features also extends to instruction level. To see how ready are DBIs to improve the transparency of analysis-environments, we provide some results in Section III-B.

Last, but not least, there is an issue regarding the checking whether a system supports these transparency attacks and to which extent. While a hard problem to formalize, at least a standard set of tests, constantly updated to latest tactics, should aid developers and users to correctly compare analysis systems from this point of view.

### *B. Problem Description*

The full range of DBI transparency problems can be summarized as follows: P1 - missing virtualization features; P2 - architectural gaps; and P3 - runtime overhead. Here is a brief description of them.

- P1. The lack of DBI virtualization for specific resources at the application or OS environment levels. This is due to the following shortcomings.
  - a) The usage of DETOUR API hooks for main OS event handling, which adds a significant overhead during the execution.
  - b) Inability to deny access to DBI address space (where DBI cache and DBI-engine resides), through standard OS API or SYS calls.
  - c) Incomplete virtualization delivered through various execution contexts. Examples of such contexts include:
    - i) CPU FPU context;
    - ii) TIB (Thread Information Block) entries like TLS (Thread Local Storage) or reserved fields used to backup Context transition variables between DBI engine and DBI cache, execution stacks, handles, etc.;
  - d) Missing environment virtualization features, responsible for revealing some of the DBI resources (e.g. OS modules enumeration APIs, OS handles, etc.), or able to fingerprint a sandbox (CPU vendor, BIOS version, specific services, etc.).
- P2. Architectural gaps, which refers to:
  - a) missing behaviors, e.g.: missing support for some assembly instructions, execution of 64 bit code in 32 bit processes, incorrect handled exceptions, etc.;
  - b) inability of the DBI to stack multiple DBI tools, in order to assist users to fill the gap between the lack of DBI virtualization and their particular analysis (this issue does not allow users to merge together the existing PIN projects to increase the DBI virtualization features).
- P3. Bad overhead management which includes:
  - a) increasing the overall allocated memory:
    - i) increasing the DBI cache size, by exploiting the fact that a DBI also instruments OS library code, the larger the number of unique API calls, the larger the cache;
    - ii) increasing the number of contexts required for threads and various instrumentation scenarios;
    - iii) spotting the increased size of an instrumented process, compared to the native execution;

TABLE I: Taxonomy-Problem relation

| Taxonomy leaf | Problem |
|---|---|
| Unsupported Assembly Instructions | P2.a |
| Unsupported Behaviors | P2.a |
| Stalling Code | P3.b |
| Memory Exhaustion | P3.a |
| Code Cache Fingerprints | P1.b |
| Instruction Pointer in Unexpected Memory Regions | P1.c |
| Incorrect Handling of Self-Modifying Code | P2.a |
| Unexpected Context | P1.c |
| Memory Region Permission Mismatches | P1.b |
| Process Hierarchy | P1.d |
| Xmode Code | P2.a |
| Incorrect Emulation of Supported Assembly Instructions | P2.a |
| Command-Line Arguments | P1.d |
| Process Handles | P1.d |
| File Handles | P1.d |
| Event Handles | P1.d |
| Shared Section Handles | P1.d |
| Signal Masks | P2.a |
| Fingerprints of DBI-related Binary Programs | P1 |
| Thread Local Storage Presence | P1.c |
| Environment Variables | P1.d |
| System Library Hooks | P1.a |
| Excessive Number of Full Access Memory Pages | P1.b |
| Common API Calls | P1.a |
| Peak Memory Usage | P3.a |
| Performance Degradation | P3.b |

b) increasing the time taken by various instrumentation behaviors (library load, branch translation, thread creation, exception handling, large loops, etc.).

### C. Contribution

Our contribution comes two fold. First, we provide a test suite comprising an extended set of state of the art transparency tests, as described in Section III. This suite covers all of the above mentioned issues and represents an important step towards creating a benchmark which may be publicly used for checking the transparency features of any analysis solution based on DBI frameworks. Analyzes of the failed tests led us to the conclusion that the main causes are given by the presence of at least one of the P1-P3 problems. The test suite was designed to extend as much as possible the taxonomy presented in [35] pg. 11, while at the same time we also provide the relation between the problems covered by the taxonomy, and the problems we have surfaced above, in Table I.

Secondly we addressed the above problems within COBAI (Complex Orchestrator for Binary Analysis and Instrumentation), a DBI engine written from scratch, specifically designed for malware analysis. A main feature of COBAI targets the DBI and OS environment transparency issues, in an attempt to successfully instrument evasive malicious binaries. As revealed by Section III, COBAI evaluates as successfully in roughly 95% of the test scenarios. The novelty of COBAI consists in a flexible and efficient combination of anti-anti-analysis techniques in order to easily obtain extendable DBI frameworks, where a main focus is providing transparency against malicious applications. While new attacks or anti-evasion techniques might surface in the future, COBAI provides an easy environment to integrate protection against them. This is sustained by the technical implementation and architecture of the core technical advances, as it allows the combination of multiple plugins ”collaborating” for obtaining better results and continuous fixes through new plugins. Even if the contribution is of incremental nature, it puts the bases for a new approach in the design of DBIs.

*Paper organisation:* The paper continues as follows: Section II provides the design and implementation details for our solution, Section III highlights the results of various evaluations also involving other state of the art solutions, Section IV presents some related work, both academically and industry oriented, and Section VI concludes our paper and offers a preview to our future work.

## II. COBAI: Design and Implementation

In this section we present an overview of the internal COBAI design, with the focus on the components assuring transparency, scalability, and analysis proliferation. The high-level design of COBAI is described in Section II-A, while the main plugins in charge with the binary analysis are introduced in Section II-C, along with their purpose and the communication process.

### A. Architecture

During our research and evaluations of other state of the art solutions, we have noticed two important and common drawbacks (validated by the evaluation, as described in Section III). First, the overall architecture of a DBI is not specifically designed to provide full virtualization and the components cannot be easily replaced. DBIs, such as Pin, usually have a closed architecture to which additional plugins may be connected. Second, critical components, such as the disassembler, API hooking mechanisms or even the DBI engine, are highly coupled. The two issues mentioned earlier are, from our point of view, deeply related, in the sense that the components that might interfere with the DBI's ability to provide transparency cannot be easily replaced without affecting the entire framework.

Therefore, developing a plugin whose sole purpose is to conceal the presence of the DBI or other system artifacts might prove useful in some cases, but will fail at tasks strictly dependent on other components and where the transparency plugin cannot interfere. The issue is, therefore, also linked to a lack of control over what the application may do, even under scenarios where tricking it is a trivial task (see Section III).

In this sense, COBAI's architecture is a modular one, where each individual component has a clear purpose and may easily be replaced or adjusted. Fig. 1a describes COBAI's architecture on a high level. The main engine is the DBI component and it has two main tasks: instrumenting the analysis and handling the plugins. The DBI loads all available plugins and, through an initialization protocol (see Figure 2), registers all APIs exposed by the plugins, which then are proxied such that any plugin may benefit from functionalities of the others, thus extending their capabilities.

A launcher is used to start both the analysis instance and a centralized analysis server by reading the analysis parameters

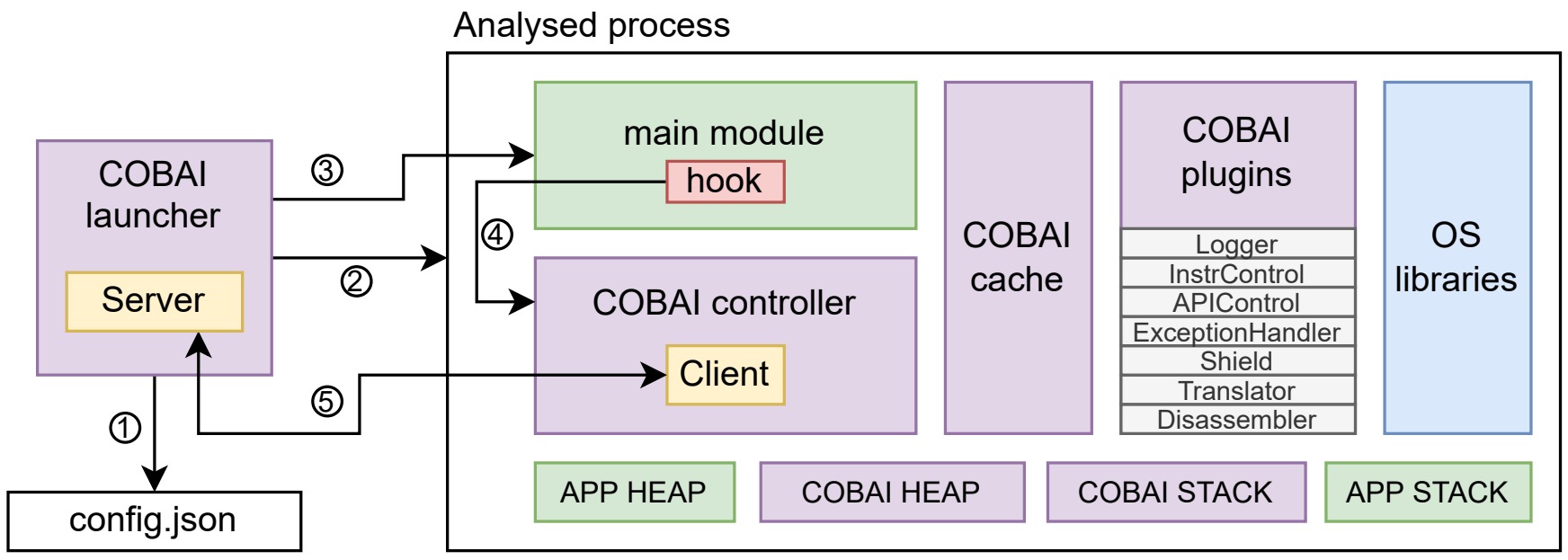

(a)

```
"paths":{
    "plugins": "C:\\COBAI",
    "target": "D:\\notepad.exe"},
"target":{
    "new_appname": "er5yop.exe",
    "debug": false,
    "use_log_folder": false
},
"plugins": {
    "plugin_names": ["disasm","translator",
        "logger","apicontrol"],
    "log_flags":  {
        "disasm": [1,2,3,4,5,6],
        "apicontrol": [1,2,3,4,5,6]},
    "plugin_param": {
        "logger": ["-1", "timer"],
        }
}
```

(b)

Fig. 1: (a) COBAI Architecture, using an enclave-like memory design. (b) Example of the config file.

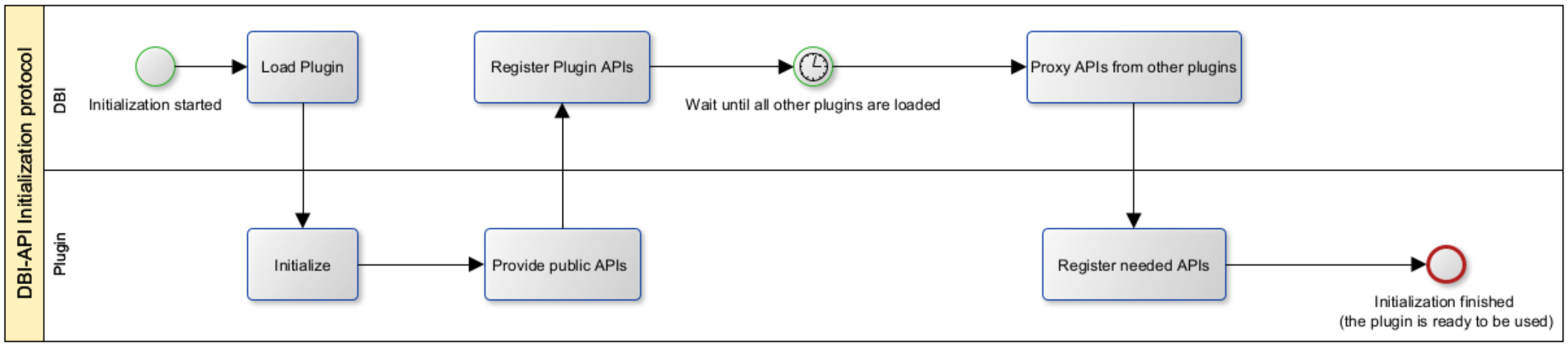

Fig. 2: DBI-Plugin Initialization protocol

from *config.json* ①. A process ② is created with the target application mentioned in the configuration file, and a payload ③ is injected into the freshly created process, further loading the COBAI controller⑤ module. The latter is responsible for loading and initializing all other dependencies (plugins). Each analysis instance registers itself through the server-client interface④. Should the analyzed application try to create a new local thread, remote thread or process (be it local or remote), the controller will announce it to the main launcher. This way, the analysis-context is extended, providing control over all available instances.

Depending on the analysis purpose and traits of interest, the user may configure COBAI by specifying a configuration file as input, using the JSON format. Fig. 1b highlights the three main sections of this configuration file. It includes information related to the analysed application (path, new name), parameters for individual plugins, logging flags and so on. Due to the development under the low coupling, high cohesion principles, the user may easily enable or disable any component, changing the behavior and the expected results.

One of the main components, "Transparency shield", has been specifically designed for ensuring transparency not only for the framework itself, but also for concealing the true nature of the environment. This is achieved by intercepting certain CPU instructions or system calls and providing "forged" results that would normally yield from a "regular" system. More details about this process are discussed in Section II-C.

The entire framework is almost self-contained, the only external dependencies being BeaEngine [34], which the disassembler plugin is based on, the Lohmann JSON library [44], used by the launcher, and Mongoose, a networking library [11]. All the libraries are statically linked, therefore the user should not worry about them.

The architectural shift to make possible the solutions mentioned above was achieved by following the low coupling high cohesion principles, by designing a flexible set of problem/issue-specific plugins, and by developing an enclave-like memory management (see Section II). The current development stage of COBAI is the result of constantly improving previous attempts and design choices. In this paper, we mainly focus on how COBAI targets the DBI and OS environment transparency issues, in an attempt to successfully instrument evasive malicious binaries.

- *availability*: while it was designed as a framework, it can be used as an out-of-the-box solution, even on highly evasive malware (see Section II-A);
- *ease of use*: no programming skills are required to produce and configure execution traces, by only adjusting a configuration file (see Section II-A);
- *development scalability*: every component is independent, having a clear and separate purpose, which leads to a dynamic architecture, such that developers may replace core functionalities with their own (including the translator or disassembler) (see Section II-C).

### B. How COBAI addresses P1-P3

P1. COBAI implements a *Shield* plugin designed to extend the resource virtualization beyond the DBI, covering a significant number of OS environment resources:

a) COBAI does not hook any of these APIs in order to capture OS events, it just captures event registration

APIs, by reserving DBI-cache address space at the registration time. The basic API hooking to capture the registration of the events is achieved through API hooks at the instrumentation level.

b) COBAI implements specific instrumentation hooks in order to slightly change the behavior of all memory-access APIs. The application under analysis is basically unable to iterate memory ranges belonging to the DBI, including DBI libraries, heap ranges, thread contexts or the DBI-cache.

c) COBAI provides virtualization for all possible execution contexts.

d) COBAI provides virtualization features for a comprehensive list of OS environment resources.

P2. COBAI is prepared to face the missing behaviors, by using a higher granularity for the DBI-engine sub-modules, thus some missing virtualization features may be added, or some plugins updated or replaced:

a) While COBAI still has some missing functionalities (execution of 64 bit code inside 32 bit process), its modular approach allows for easy localization of incomplete or wrong behaviors in order to implement the missing functionality. For instance, any missing ASM instruction should be solved by either replacing the entire disassembler module or perform an update of the module.

b) COBAI is designed to layer the execution of different DBI-tools at the same time. In this regard, it is able to execute the *Shield* plugin and the instruction and API tracing plugins all at the same time.

P3. COBAI implements a minimum of overhead management, such that while not fully able to make it disappear, the measurements are adjusted such that it gets apparently lower compared to the actual one:

a) While still under development, COBAI may face this challenge by moving large DBI-cache portions outside the process address space, if this limit is pushed behind the edges.

b) COBAI is able to overcome some of these time overheads through a carefully sync between CPU counters and execution context counters or time-specific APIs. While some applications indeed take longer to instrument, the application is unable to determine that the time overhead is grater compared to a native execution.

### C. Plugins

By design, each feature or group of features is handled by a specific plugin or a small set of dependent plugins, in order to be easily updated or even replaced. The core set of plugins is presented in Fig. 3, where *COBAI Controller* itself is a plugin.

While the modularity should provide decoupling, *Controller* along with *Translator* and *DISASM* plugins are mandatory for a proper binary translation, but a basic analysis may be performed without the other plugins. The main idea, in this case, is that the user may customise the usage of COBAI according to countless analysis scenarios, thus increasing its efficiency: track only the executed instructions or API calls, enable or disable the handling of polymorphic applications, handle exceptions, provide transparency and so on. Another important remark is that, while plugins are natively independent, some of their features rely on the presence of others. For instance, the plugin providing transparency requires the existence of plugins handling APIs or CPU instructions. The dependency relations are included in the plugins' description.

Following, we briefly describe the plugins currently implemented in COBAI and describe scenarios where they are needed:

*COBAI Controller::* This plugin, along with the *Translator* and *DISASM*, are the minimum requirements for COBAI to operate. This plugin itself provides support for: the management of other plugins; API interface between plugins; instantiation of a new analysis (be it a different process, a different thread inside or outside the process under analysis, or a callback or exception handler).

*Translator::* The flow of an actual instrumentation of the target code is found in *Translator* module. This plugin is responsible with code translation, cache management, and thread state management. Compared to other tools and frameworks where the translator is part of the DBI, in COBAI it is possible to totally replace the translation mechanics and all its underlying aspects, with a totally different approach, either for supporting additional architectures, for performance or for academic reasons.

*DISASM::* This plugin handles the disassembly of instructions and it is currently an interface for the BeaEngine disassembler. It is mainly used by the *Translator* plugin and also provides the string version of instructions, in the NASM Intel syntax. Replacing the disassembler also requires a specific interface to bind to *Translator* module.

*ExceptionHandler::* This plugin provides support for handling exceptions and also aids in translating exception filters and handlers. It supports the following exceptions: Win32 Structured Exception Handling (SEH); Vectored Exception Handling (VEH); Unhandled Exceptions; C++ exceptions.

*APIControl::* This plugin assists in:

- API detection, along with providing details regarding the API name, parameters, and library;
- registration of user-callbacks to be called before and/or after a specific API, along with the option to skip the actual execution of the function or simulate its behaviour, correctly adjusting the stack and other necessary parameters.

Instead of relying on endless lists of symbols (such as PDB files) provided by vendors (e.g. Microsoft), the plugins makes use of a lightweight binary structure providing various details, such as API names, parameters, their type and so on. The binary structure is inspired by the *apis_def* component presented in [20], which is a plugin for x64dbg [19].

*InstrControl::* This plugin assists, similarly to the APIControl, in the registration of user-callbacks which are to be called before and/or after an instruction or group of instructions. The group refers to the fact that a plugin may register a callback not only for a specific instruction (referenced by opcode, for instance), but also by a regex, instruction type (control, transfer, branching, etc) or any combination of the above. The operations are handled on a binary level, such that the overhead is minimal. Also, any registered callback has the possibility of manipulating the instruction or specifying that it should not be executed.

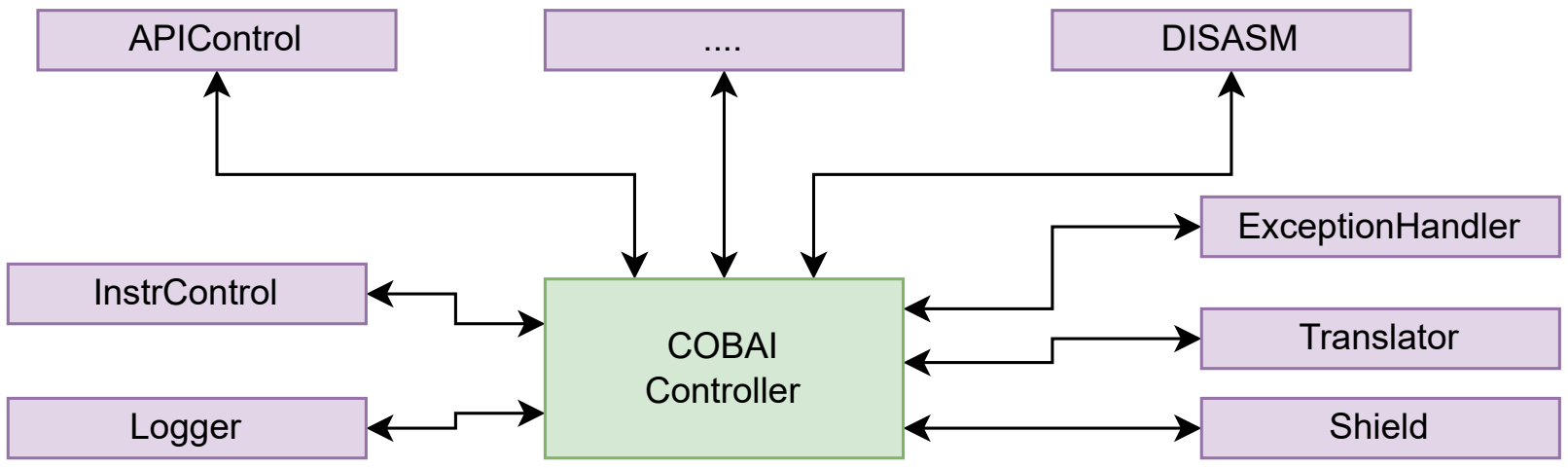

Fig. 3: Plugin Infrastructure

*Logger::* The main role of this plugin is to facilitate logging capabilities for other plugins. It allows the following: registration of callbacks when needed; creation of new trace logs; multithreaded support; logging any type of output (this is also helpful for plugins needing to dump memory regions); cache support, increasing the performance of the overall analysis.

*Shield::* This plugin is responsible with transparency policies both for the DBI engine and for the OS environment. The transparency is mainly achieved through various routines registered as callbacks, using the *APIControl* and *InstrControl* plugins. Therefore, the plugins ensures that forged results are passed to instructions or API (system) calls for events that would trigger details about the environment. For instance, fake results are provided to instructions or APIs querying the presence of a VM context (VM features are present, files or registry corresponding to several VM hypervisors or emulators, time difference between instructions and so on).

These routines are also grouped, such that a user may enable or disable sets of routines, according to the scenario (file manipulation, system queries, registry access, network connections, etc). Additionally, this plugin partially enhances the DBI address space, such that it behaves similarly to a trusted execution environment (TEE) / enclave. This ensures that the analysed application has no access to the DBI memory. Furthermore, any instrumented child/remote process will be forbidden to access the instrumented parent DBI address space.

## III. Evaluation

This section evaluates the virtualization features of COBAI along with four other DBI frameworks (PIN, DynamoRIO, FLOW, QBDI), considering two points of view: the virtualization rate of the DBI framework itself facing specific DBI-transparency attacks in Section III-A, and the virtualization of the analysis environment resources (e.g.: Operating System, Sandbox - where the DBI executes) facing specific Sandbox fingerprints in Section III-B.

To evaluate the transparency of the DBI, the following ingredients are required:

1) an analysis environment as the running context for the DBI;
2) a DBI-engine (back-end of the analysis, mandatory);
3) a DBI-tool running on top of the DBI-engine, capable to fill the virtualization gap;
4) a target application executing certain transparency attacks (either for the DBI-engine or for the Sandbox environment).

These ingredients are configured differently across Sections III-A and III-B, to capture a series of problems we have identified during our experiments.

### A. DBI Transparency Evaluation

This experiment highlights the transparency features of COBAI, compared with several other DBI frameworks. In some of our experiments, we have used specific DBI-tools (to assist the mitigation of the transparency issues) that were publicly available, while in others we only used the DBI-engine itself performing the basic binary instrumentation. The motivation for the chosen configuration for DBI-tools is presented in more details in Section III-A2, while sections III-A1 and III-A3 describe the test-suite we have used to perform the transparency attacks and the results we obtained, respectively.

*1) Test suite:* The test suite consists of 57 applications. A part of them were collected from other research projects, while the rest were either inspired from the informal description found in the literature, or derived from our experience. As a consequence, we expanded the initial set of tests. Therefore, our test suite has been designed and developed based on four groups of public available proof of concepts, while the last two groups extend some uncovered scenarios:

1) Publicly available tests developed for Windows OS:
   - 12 test created by AVAST in 2014 for [38];
   - 1 test for a timing attack described in [49];
   - 4 tests described in [29].
2) 12 publicly available applications developed for other OS (e.g.: Linux), but adapted by us to execute on Windows described in [59], the remaining one (12 out of 13) is Linux-specific and could not be ported to Windows OS because of the differences in the OS kernels.
3) 14 tests based on informal descriptions presented in [35] and partially on source-code fragments described in [43], [42], [36], [39].
4) 14 tests designed by us, implementing transparency attacks as an extension of the previous group of tests.

*A brief description of the tests::* The name of the tests can be found in the second column of Table III. The tests are split into six groups based on their source. The first four groups of tests were public available [38], [49], [59], [29]. The fifth set of tests was developed based on informal descriptions and partial source code found in the references of [35]. The sixth group has been fully developed and implemented through our research, as the attack scenarios were created from scratch or extending/deriving from others.

The first [38] and fourth [29], [15] sets of tests, include experiments in which the memory access (code, heap, stack, execution contexts) is manipulated, triggering various exceptions in order to expose the lack of DBI resources virtualization, or

TABLE II: Extended evaluation for DBI transparency issues
**AR**: Arancino (x86 on Win10 x64) for PIN 2.14; **VMS**: PinVMShield (x86 on Win10 x64) for PIN 3.20; **SoK**: SoK (x86 on Win10 x64) mitigations for PIN 3.20; **PIN**: stand-alone PIN (x86 on Win10 x64) 3.20; **BP10**: BluePill (x86 on Win10 x64) for PIN 3.20; **BP7**: BluePill (x86 on Win7 x64) for PIN 3.16; **JLP**: JuanLesPIN for PIN 3.20;**DR**: stand-alone DynamoRIO (x86 on Win10 x64); **QBDI** (x86 on Win10 x64) 0.8.0; **FLOW** (x64 on Win10 x64)

| Ref | Test Name | COBAI | AR | VMS | SoK | PIN | BP10 | BP7 | JLP | DR | QBDI | FLOW |
|---|---|---|---|---|---|---|---|---|---|---|---|---|
| [38] | ExecuteData1 | ✓ | ✗ | ✓ | ✓ | ✓ | ✓ | ✗ | ✓ | ✗ | ✘ | ✘ |
| | ExecuteData2 | ✓ | ✗ | ✓ | ✓ | ✓ | ✓ | ✗ | ✓ | ✗ | ✘ | ✘ |
| | ExecuteUnmap1 | ✓ | ✗ | ✓ | ✓ | ✓ | ✓ | ✓ | ✓ | ✗ | ✘ | ✘ |
| | ExecuteUnmap2 | ✓ | ✗ | ✗ | ✗ | ✗ | ✗ | ✗ | ✓ | ✗ | ✘ | ✘ |
| | ExecuteUnmap3 | ✓ | ✓ | ✓ | ✓ | ✓ | ✓ | ✓ | ✓ | ✓ | ✘ | ✘ |
| | ExecuteUnmap4 | ✓ | ✓ | ✓ | ✓ | ✓ | ✓ | ✓ | ✓ | ✓ | ✘ | ✘ |
| | ExecuteUnmap5 | ✓ | ✘ | ✓ | ✓ | ✓ | ✓ | ✓ | ✓ | ✗ | ✘ | ✘ |
| | FpuContext1 | ✓ | ✗ | ✗ | ✓ | ✗ | ✓ | ✓ | ✓ | ✓ | ✗ | ✗ |
| | FpuContext2 | ✓ | ✗ | ✗ | ✗ | ✗ | ✗ | ✗ | ✗ | ✗ | ✗ | ✗ |
| | ServiceException1 | ✓ | ✓ | ✘ | ✘ | ✘ | ✘ | ✘ | ✘ | ✗ | ✘ | ✓ |
| | TransientException1 | ✓ | ✗ | ✗ | ✓ | ✗ | ✗ | ✗ | ✓ | ✘ | ✘ | ✘ |
| | TransientException2 | ✓ | ✗ | ✗ | ✗ | ✗ | ✗ | ✗ | ✗ | ✓ | ✘ | ✘ |
| [49] | Detector | ✓ | ✗ | ✗ | ✗ | ✗ | ✗ | ✗ | ✗ | ✓ | ✓ | ✓ |
| [59] | enter | ✓ | ✓ | ✓ | ✓ | ✓ | ✓ | ✓ | ✓ | ✓ | ✓ | ✘ |
| | envvar/strings | ✓ | ✓ | ✘ | ✗ | ✗ | ✗ | ✗ | ✗ | ✗ | ✓ | ✓ |
| | find-constant | ✓ | ✗ | ✗ | ✗ | ✗ | ✗ | ✗ | ✓ | ✓ | ✓ | ✗ |
| | jit-branch | ✓ | ✗ | ✗ | ✗ | ✗ | ✓ | ✓ | ✗ | ✗ | ✗ | ✓ |
| | jit-lib | ✓ | ✓ | ✘ | ✓ | ✓ | ✓ | ✓ | ✘ | ✓ | ✓ | ✓ |
| | mapname | ✓ | ✗ | ✗ | ✗ | ✗ | ✓ | ✗ | ✓ | ✗ | ✗ | ✗ |
| | NX | ✓ | ✗ | ✓ | ✓ | ✓ | ✓ | ✗ | ✓ | ✗ | ✗ | ✗ |
| | PagePerm | ✓ | ✗ | ✗ | ✗ | ✗ | ✗ | ✗ | ✗ | ✗ | ✓ | ✗ |
| | FXSAVE | ✓ | ✗ | ✗ | ✗ | ✗ | ✗ | ✗ | ✗ | ✓ | ✗ | ✗ |
| | IPSIGINFO | ✓ | ✗ | ✗ | ✗ | ✗ | ✗ | ✗ | ✗ | ✗ | ✘ | ✗ |
| | SMC | ✓ | ✓ | ✓ | ✓ | ✓ | ✗ | ✗ | ✗ | ✓ | ✗ | ✓ |
| | VMLeave | ✓ | ✗ | ✗ | ✗ | ✗ | ✗ | ✓ | ✘ | ✗ | ✘ | ✘ |
| [29] | PageGuard Single | ✓ | ✗ | ✗ | ✓ | ✗ | ✗ | ✗ | ✓ | ✓ | ✘ | ✘ |
| | PageGuard Multi | ✓ | ✗ | ✗ | ✓ | ✗ | ✗ | ✗ | ✗ | ✗ | ✘ | ✘ |
| | read CC | ✓ | ✓ | ✗ | ✓ | ✗ | ✗ | ✗ | ✓ | ✓ | ✓ | ✘ |
| | FPU | ✓ | ✗ | ✗ | ✓ | ✗ | ✓ | ✓ | ✓ | ✓ | ✗ | ✗ |
| [35] | ProcessHirarchy | ✓ | ✗ | ✗ | ✗ | ✗ | ✗ | ✗ | ✗ | ✗ | ✓ | ✗ |
| | cmdlnargs | ✓ | ✗ | ✗ | ✗ | ✗ | ✗ | ✗ | ✗ | ✗ | ✓ | ✓ |
| | ProcessHandles | ✓ | ✓ | ✘ | ✓ | ✓ | ✓ | ✓ | ✓ | ✓ | ✓ | ✓ |
| | FileHandles | ✓ | ✗ | ✘ | ✗ | ✗ | ✓ | ✓ | ✗ | ✓ | ✓ | ✓ |
| | SectionHandles | ✓ | ✓ | ✓ | ✓ | ✗ | ✓ | ✓ | ✗ | ✓ | ✓ | ✓ |
| | LibraryHooks | ✓ | ✓ | ✓ | ✗ | ✗ | ✗ | ✗ | ✗ | ✗ | ✓ | ✓ |
| | HeapStack | ✓ | ✓ | ✓ | ✓ | ✓ | ✓ | ✓ | ✓ | ✓ | ✓ | ✓ |
| | MemExhaus | ✓ | ✗ | ✗ | ✗ | ✗ | ✗ | ✗ | ✗ | ✗ | ✗ | ✘ |
| | Xmode | ✘ | ✘ | ✘ | ✘ | ✘ | ✘ | ✘ | ✘ | ✘ | ✘ | ✘ |
| | TLSPresence | ✓ | ✗ | ✗ | ✗ | ✗ | ✗ | ✗ | ✗ | ✗ | ✓ | ✗ |
| | APIMonitor | ✓ | ✗ | ✓ | ✓ | ✗ | ✓ | ✘ | ✗ | ✗ | ✗ | ✓ |
| | StallingCode | ✗ | ✗ | ✗ | ✗ | ✗ | ✗ | ✗ | ✗ | ✗ | ✗ | ✗ |
| | escape_dbi | ✓ | ✗ | ✗ | ✗ | ✗ | ✗ | ✗ | ✗ | ✗ | ✗ | ✗ |
| | peak_mem_usage | ✓ | ✗ | ✗ | ✗ | ✗ | ✗ | ✗ | ✗ | ✗ | ✗ | ✗ |
| | unhandled_instrs | ✓ | ✘ | ✗ | ✗ | ✘ | ✗ | ✗ | ✗ | ✓ | ✘ | ✘ |
| III-A1 | exception_mix | ✓ | ✓ | ✓ | ✓ | ✓ | ✓ | ✓ | ✓ | ✓ | ✘ | ✘ |
| | stacktrace | ✓ | ✘ | ✗ | ✓ | ✓ | ✓ | ✓ | ✓ | ✓ | ✘ | ✓ |
| | additional_threads | ✓ | ✓ | ✓ | ✓ | ✓ | ✓ | ✓ | ✓ | ✓ | ✓ | ✓ |
| | additional_stacks | ✓ | ✗ | ✗ | ✗ | ✗ | ✗ | ✓ | ✗ | ✘ | ✗ | ✓ |
| | free_unknown | ✓ | ✘ | ✘ | ✘ | ✘ | ✘ | ✘ | ✘ | ✘ | ✘ | ✘ |
| | API_unhook | ✓ | ✓ | ✗ | ✘ | ✘ | ✘ | ✗ | ✘ | ✘ | ✘ | ✓ |
| | abuse_exceptions | ✓ | ✘ | ✘ | ✘ | ✘ | ✘ | ✘ | ✘ | ✘ | ✘ | ✘ |
| | Transit_x64 | ✘ | ✘ | ✘ | ✘ | ✘ | ✘ | ✘ | ✘ | ✘ | ✘ | ✘ |
| | ChildFeedback | ✓ | ✗ | ✗ | ✗ | ✗ | ✗ | ✗ | ✗ | ✗ | ✓ | ✓ |
| | ThreadExhaustion | ✓ | ✘ | ✘ | ✘ | ✘ | ✘ | ✘ | ✘ | ✗ | ✓ | ✘ |
| | ThreadContext | ✓ | ✓ | ✓ | ✓ | ✓ | ✓ | ✓ | ✓ | ✘ | ✓ | ✓ |
| | thread-increase-DBI-cache | ✓ | ✘ | ✘ | ✘ | ✘ | ✘ | ✘ | ✘ | ✘ | ✓ | ✘ |
| | API-calls-increase-DBI-cache | ✓ | ✘ | ✘ | ✓ | ✓ | ✘ | ✘ | ✘ | ✓ | ✓ | ✘ |

✓ - test passed ✗- test failed ✘ - instrumentation crashed to lead to unexpected behavior. These tests expose various faces of problem P1. The second group consisting of a single test [49] describes an application performing a timing attack. The difference is an overhead visible during the analysis (this test is related to problem P3). The third [59], fifth [35] and the last sets of tests (designed by us) include experiments exposing a wide range of faces for problems P1, P2, P3. Overall, our test-suite covers the DBI transaprency problems as follows:

- P2 - architectural gaps: six tests (*enter, abuse_exceptions, exception_mix, Transit_x64, unhandled_instrs, Xmode*);
- P3 - increased runtime overhead: ten tests (jit_branch, API_Calls_Increase_DBI_cache, peak_mem_usage, StallingCode, thread_increase_DBI_cache, MemExhaus, abuse_exceptions, jit_lib, Detector, ThreadExhaustion);
- P1 - missing virtualization features: 41 tests (all the others).

The 41 tests triggering P1 problem, expose various common lacks of virtualization. While they could have been also passed by any other DBI framework, it seems that none of the DBI frameworks / DBI tools was able to fully pass them. COBAI is capable to stack various plugins as DBI tools and execute them at once, addressing this way the P2.b problem. Below we highlight home some interesting tests are capable to triggering P2.a and P3 problems:

- P2.a-*Transit_x64*: While all the tested applications agree with a uniform execution with respect to the target architecture of a process, this test executes both 32 and 64 bit code in the same process. The execution to 64 bit code is possible by making a far call using the 64 bit code segment (already present in 32 bit applications - also known as Haven's Gate [8]), to a 64 bit piece of code and then back to 32 bit using the 32 bit code segment. The *Xmode* test, extends this approach and leverages the WOW gateway to go even further and make API calls to legitimate 64-bit APIs in a 32 bit process.
- P2.a-*abuse_exceptions*: All the DBIs implement some minimum support for exception handling, however, we have noticed that abusing the exceptions mechanisms leads to crashes and sometimes to executions refusing to finish.
- P3-*Detector*: Test causing significant performance issues, creating a behavioral difference between the CPU cache and the DBI cache, when dealing with thread synchronization.
- P3-*StallingCode*: Test leading to a significant performance degradation compared to native execution (the difference exceeds 1000%).

*2) Evaluation Process:* We executed all the test applications in a Sandbox-like environment running on top of VMWare 16 with hardware virtualization enabled. The operating system inside the virtual machine was a Windows 10 x64 for most of the tests (or Windows 7 x64 for PIN 3.16 - an exception test) with an i9-10885H CPU running at 2.4GHz. The virtual machine had 2 cores (4 threads), and a total of 4GB of memory.

We used other four DBIs to compare to COBAI: *Intel PIN* [10], [45] versions: 3.20, 3.16, 2.14, DynamoRIO [25], [6] version: 9.0.18983, QBDI [37] version: 0.8.0, FLOW [7]. Because COBAI, at the moment, is built for the x86 architecture, three (PIN, DynamoRIO, QBDI) of the DBIs were also used under a similar context. FLOW, on the other hand, was configured for Windows x64, as it only supports this architecture. Given the differences between the DBIs, we built the tests both for Windows x86 and Windows 64, operating the low-level attack differences to behave the same. Beside the two different versions of the tests source-code (Windows x86, Windows x64), we also had to build a third version of the tests, to execute on QBDI as the instrumentation in this case is performed for a function and not for a standalone application.

We found DBI-tools only for PIN, this is the reason for which we also considered PIN with no tool at all running on top. The motivation for also using the standalone version of PIN is motivated by our will to compare the increase of fixed transparency issues provided by the tools, with the results of not using a tool at all. For COBAI we have used the Shield plugin presented in Section II, while for DynamoRIO, FLOW and QBDI we only used the DBI-engine itself. As PIN has a lot of versions and some of them are pretty different, we configured the PIN-tools and PIN versions as follows:

- for PIN 2.14 we used Arancino [50], [3];
- for PIN 3.16 we used BluePill [30], [4];
- for PIN 3.20 we used PinVMShield [51], [13], SoK [29], [15], the PIN DBI-engine itself without a specific tool, and BluePill;

The set of tests was executed using the required architecture for each DBI (x64 for FLOW and x86 for all the others). The following labels were used to tag the possible results in Fig. 4:

- the DBI was not detected (TEST PASSED);
- the DBI was detected (TEST FAILED);
- the execution generated an application crash as a result of the evasion mechanism being instrumented (EXEC ERROR).

None of the tests generate any crash by themselves (i.e. without being instrumented), ensuring the fact that any generated crash, during the instrumentation, was caused by the DBI-engine or the logic inside the DBI-tool. For BluePill we used both PIN 3.16 (as suggested by the developers - where Windows 7 was also a requirement) and PIN 3.20 with additional changes required to make possible the build process.

*3) Results:* The results may be interpreted from two points of view. The first one is a statistical overview of the results in Fig. 4, while the second one is a discussion on side-effects of implementing or lacking features to assist problems P1,P2,P3.

At the highest level, Fig. 4 may be interpreted as follows:

- Overall COBAI passed $\approx$ 95% of all tests and missed three tests where everyone else, also failed;
- PIN passed $\approx$ 34% on average and DynamoRIO $\approx$ 38%; as the results show, DynamoRIO has an increased $\approx$ 4% for passed tests compared to PIN, and the same rate for lower failed tests; these differences tells us that DynamoRIO performs better compared to PIN;
- QBDI and FLOW scored both 61%-66% for the failed tests, having the same number of ERROR tests; the passed tests are also close for these DBIs ($\approx$37% for QBDI and $\approx$33% for FLOW).

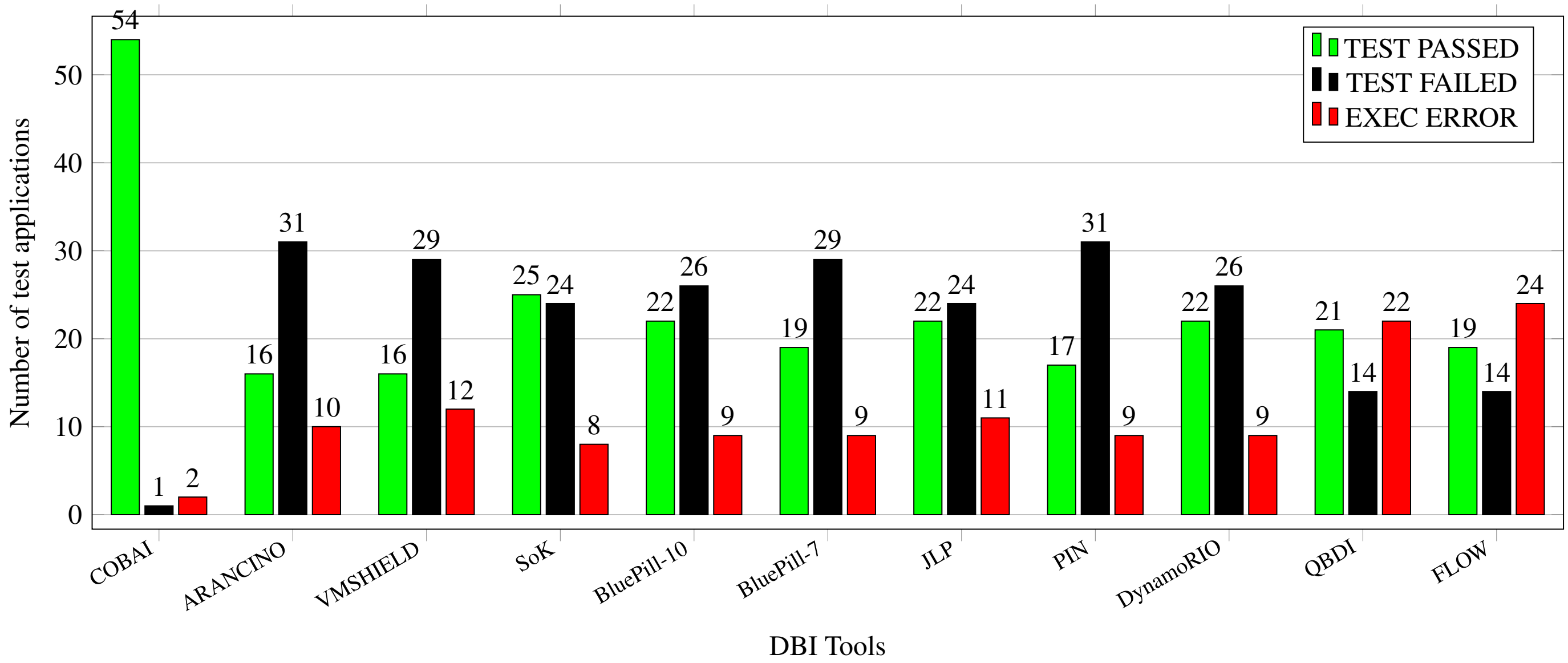

Fig. 4: DBI Transparency across a total of 57 applications and a range of 5 (COBAI, PIN, DynamoRIO, QBDI, FLOW) different DBIs, where PIN had configured 5 different PIN tools (Arancino, VMShield, SoK, BluePill, JuanLesPin) as well as the standalone version (PIN).

Below is a short discussion of the side-effects generated by the presence or the lack of virtualization features leading to problems P1-P3:

*P1 - Full virtualization coverage for own DBI resources:* Our set of tests highlights the fact that this problem is a general one. None of the DBIs approach in a serious way to this subject, leaving DBI-tools to fix this lack of virtualization for the moment (and possible forever). This virtualization gap is possible because the developers of these analysis frameworks and their users, do not agree the same on how far the virtualization of the DBI resources should go. In this regard, most of the DBI engines only cover some basic traits, such as context switching, or strict behavioral changes for event handling. From our experiments, none of the existing DBIs attempt to fully isolate themselves. Even the border-line isolation is not bullet-proof. Tests **APIMonitor, VMLeave, IPSIGINFO** described in Section III-A1 confirm our statement. From our point of view, to increase the security of the DBI at this level, would require to fully isolate the DBI resources. COBAI creates a border between the DBI resources and the instrumented application, and isolates itself by slightly shifting the application behavior on queries targeting any of the DBI resources.

*P2 - Architectural gaps:* While this is the most challenging problem to solve, COBAI already incorporates some features ready to assist upcoming issues in this area. The engine of the DBI uses modularity in a way that core sub-components can be individually replaced or updated. COBAI is able to pass 4 out of 6 tests revealing this problem and the other 2 are work in progress, whereas all the other DBIs may take more to add the missing functionalities. A lateral face of the COBAIs architecture and possible one of the reasons for which we were able to obtain the highest score, is the possibility to execute multiple DBI tools at once. To support this approach, a DBI must have its architecture designed such that multiple DBI-tools hooking the same instructions would not cause any conflict. The other DBIs are required to merge together different DBI-transparency mitigation rules, or even merge a custom logic to perform a specific analysis with the focus on fixing the same time some virtualization gaps. This feature is not something that is visible in our evaluation but has an indirect impact. During the development of COBAI we have used both the Shield plugin and a tracing tool,allowing us to understand and solve most of the issues we have faced in a timely fashion.

*P3 - Runtime overhead:* This problem affects the ability of the DBI to provide the desired results by significant performance degradation. This is an indirect approach to reveal the presence of the DBI, as it does not rely on DBI artifacts or fingerprints, but on critical differences in execution time, or resource usage. **StallingCode** and **Detector** described in Section III-A1 provide concrete examples for such scenario. To be able to perform well under such circumstances, a DBI should mix runtime optimizations and behavior shifts in time measurement.

## B. *Sandbox/Environment Transparency Evaluation*

The second experiment highlights the capacity of a DBI framework to conceal the true nature of the environment, making it look like a system of a regular user. In other words, the analysed application might look for certain artifacts, such as files, processes, registry keys, resources usage and so on, causing the environment to be detected as a Virtual Machine, Sandbox, Analysis environment, etc. This draws attention to the importance of concealing any environment artifacts that reveal its true purpose. This experiment is related to the previous one, but with additional tweaking of the virtualization support behind the OS, and we also reduced the number of tests to those implementing explicit sandbox evasion techniques.

*1) Test suite:* For this evaluation we have found two publicly available projects (pafish [12] and al-khaser [1]) that together implement more than 300 sandbox fingerprinting tests. The tests are classified based on a common purpose (i.e.: detection

TABLE III: Environment transparency for virtual machines; the lower the values, the more transparent the environment becomes to the instrumented application. X represents an application crash.
Legend: AR - Arrancino for PIN (ver. 2.14); VMS - PinVMShield for PIN (ver. 3.20); SoK for PIN (ver. 3.20) - [29], [15]; PIN - stand-alone PIN (ver. 3.20); BP - BluePill for PIN (ver. 3.20); DR - DynamoRIO (ver. 9.0.1898)

| VM | Sample | COBAI | AR | VMS | SoK | PIN | BP | DR | QBDI | FLOW | Native |
|---|---|---|---|---|---|---|---|---|---|---|---|
| VMWare | pafish | 0 | 10 | X | 10 | 10 | X | 11 | 10 | 11 | 11 |
| | al-khaser | 1 | X | X | X | X | X | 45 | X | 49 | 44 |
| VirtualBox | pafish | 0 | X | X | 20 | 21 | X | 21 | 21 | X | 22 |
| | al-khaser | 1 | X | X | X | X | X | 79 | X | 83 | 78 |
| Hyper-V | pafish | 0 | X | X | 5 | 5 | X | 5 | 6 | X | 5 |
| | al-khaser | 1 | X | X | X | X | X | 33 | X | X | 29 |
| QEMU-KVM | pafish | 0 | X | X | 5 | 5 | X | X | 6 | X | 5 |
| | al-khaser | 1 | X | X | X | X | X | X | X | 39 | 30 |

Total number of available tests for al-khaser: 283 Total number of available tests for pafish: 55

of virtual machines like VMWare [17] or VirtualBox [16], detection of debuggers, detection of injected modules, timing attacks, etc.). Other public projects (*unprotect* [21] and *evasions checkpoint* [41]) with the same target (exposing specific sandbox and analysis environments fingerprints), share a lot of code from the two selected applications.

*2) Evaluation Process:* This evaluation is based on executing the two applications (*pafish 6.2* and *al-khaser 0.80*) on four different virtual machines, with support for hardware virtualization (Intel VMX in our case): VMWare Workstation 16.1.1, VirtualBox 5.2.24, Hyper-V and KVM using QEMU 2.5.0. The configuration of the Sandbox was identical with that used for the previous evaluation, in Section III-A2. Table III highlights our results. We used a red cross where the execution crashed, or the total number of triggered sandbox artifacts. The lower the number, the better.

*3) Results:* The obtained results are included in Table III. Each line in the table contains the total number of detection tests for the specific DBI, out of a maximum of 283 on *al-khaser* and 55 on *pafish*. This experiment highlights the fact that, while COBAI was able to pass all the test scenarios, other DBIs either kept crashing (Arancino, VMShield, BluePill) or triggered a significant number of detections. Some of the DBIs have also triggered an increased number of detections compared to a native execution of the test (no instrumentation). This is because the DBI itself introduced aspects that triggered additional detections, even though they do not directly check the presence of a DBI framework, but rather of an analysis solution in general. For instance, Flow triggered an increased number of detections for VirtualBox (by 6%) and KVM (by 30%), while DynamoRIO has a significant number of total detections for Hyper-V (by 13%). The only detection triggered by COBAI may be considered as a false positive, since the test involves the detection of how the binary code is aligned, but it is meaningless as long as the application is compiled with size optimizations. We have also expected BluePill to provide better results for this experiment, since it is in a more advanced stage towards transparency compared to its corresponding PIN DBI tools (Arancino, SoK, PinVMShield). However, we found that it constantly triggered application crashes for *pafish*, due to Windows10 WMI (Windows Management Instrumentation) usage, and for *al-khaser* due to incorrect exception handling. Testing the instrumentation of *al-khaser* on all versions of PIN, up to 3.22, we found that most probable there is an architectural issue, as PIN kept crashing without using a specific tool.

### C. Overall discussion

From our evaluation, we conclude that existing DBIs are not yet ready to virtualize a wide range of OS resources. This lack of virtualization makes them unsuitable for automating the analysis of malicious binaries. While other analysis environments may also share some of these transparency issues and lack of resource virtualization, we believe that DBIs may be the first analysis environments to provide good results in this regard. COBAI provides reliable solutions to problems P1,P2,P3, with some caveats related to P2 and P3. We draw attention again to the architecture, as these issues are difficult to solve for other DBI frameworks, since they might require a complete redesign. Moreover, the sandbox related issues only seem to add more challenges to the already existing ones, related to the DBI itself.

## IV. Related Work

Generally, DBIs may assist the analysis of applications with consistent data collected at runtime, that may further be used for code profiling, statistics or behavior extraction. Our use-cases target behavior extraction in malicious binaries performing analysis-evasion scenarios. For our evaluation we have used DBI frameworks like PIN [45], DynamoRIO [25], QBDI [37] and FLOW [7]. In last decade, other analysis tools developed on top of these frameworks, some of them targeting benign binaries, while others targeting malicious applications. Because it is pointless to talk about the transparency of these frameworks dealing with benign binaries, we came across some projects and research works that also deal with this attack surface. TRITON [18] is a framework based on PIN, capable of performing concolic execution. While it does not target malicious binaries in a specific way, we know that existing PIN-tools attempting to fix PIN transparency issues, may be mixed with it, to perform concolic execution for APTs (Advanced Persistent Threats), for instance. Other DBI-based analysis tools like BAP [26] and ANGR [53] are able to analyze malicious binaries or payloads up to a certain level, but they do not attempt to overcome the analysis evasion as we would expect to.

The benefits of using DBI frameworks for malware analysis were explored by many research projects, and, as a consequence, the general interest and demand has risen. For instance, tools and frameworks like Pindemonium [28], RePEConstruct [40],

Arancino [50], PinVMShield [13], BluePill [30] and PEMU [58] provided good results at certain specific tasks, but the lack of a higher purpose, transformed them into proof of concepts for possible other projects. Other types of issues have been uncovered targeting DBI frameworks. For example, in Section E of a similar research [42], authors identify four distinct DBI evasion mechanisms found in commercial protectors/packers which allows applications to escape the analysis environment. Considering all these premises, DBIs seem to reach a dead-end, which might explain the reluctance of security products or analysis environments to incorporate such techniques. Our purpose was to overcome this disadvantages and change the current mindset, thus developing COBAI.

As an alternative to binary instrumentation, researchers have been replacing it with emulation introspection like QEMU [24], hypervisor introspection found in [55], [54], [33], or API hooking found in tools like FRIDA [48] and the monitor component of Cuckoo Sandbox [5]. During our research, we came across various research projects (be them theoretical or practical) describing specific DBI or Sandbox transparency attacks, or providing transparency-fixes for specific DBI frameworks:

- SafeMachine [38] is a set of 12 demos proposed by AVAST, to attack the transparency of DBIs through exceptions; most the demos involve exceptions to leak DBI specific content (e.g.: DBI stack, DBI cache, DBI memory management);
- SoK [29] is a research proposing four transparency attack scenarios as well as their countermeasures for PIN;
- PwIN [59] is a research discussing the transparency issues of Intel PIN on Linux; the authors implement a set of about 13 demos to state their point and include the most variate set of attacks, including some that are able to detect the DBI execution cache, and even escape the analysis;
- Detector [49] is a research proposed by the same author of the FLOW DBI; the demo exploit a low-level CPU-cache behavior taking place in a thread-sync mechanism;
- Some classifications for the DBI transparency issues in [35] and [31]; we took the informal descriptions and references found in these papers to implement 13 demos for our evaluation;
- Code examples for concrete DBI transparency issues in [42], [39], [43], [36], [23];
- A series of DBI tools like Arancino [50], BluePill [30], PinVMShield [51], ProcTracer [14], RePEConstruct [40];
- Sandbox transparency attacks in al-khaser [1], pafish [12] and some public available sources [41], [21] citing most of the examples from the two projects.

All these works draw special attention to how DBIs are able to virtualize specific execution-context resources like certain CPU instructions, API functions, memory access, callbacks and exceptions, execution time, etc. The aspects handled by the authors cover: leaking DBI internal state, transparency, performance, escaping instrumentation etc. We considered all the possible DBI-transparency issues presented in these research papers as a whole, and even extended it with new scenarios in an attempt to create a reference benchmark where COBAI for the moment is on top of all DBIs we have tested.

## V. Limitations and Future Work

COBAI has been designed with a flexible architecture that facilitates its extension with new features and functionalities. Because COBAI is still under development, it is currently limited to Windows 32-bit applications. Some other limits are the inability to instrument x64 code on x86 processes, and lacking some runtime optimizations for overhead management. However, the overcoming og these limits are merely a time problem. While COBAI successfully overcomes multiple others analysis challenges, we developed a plan for future improvements, separated into categories, based on the purpose and type:

*DBI Transparency Benchmark Improving::* Starting from the current state of our work, we plan to extend the set of tests both for DBIs and also for the environment and make it publicly available. The benchmark we proposed will be updated such that it would become a measuring instrument for DBIs promising transparency and we also hope to formalize and automate the definitions for transparency-attacks as well as testing a certain DBI or sandbox environment against a specific transparency-attack.

*Performance and soundness of instrumentation::* COBAI will be extended to support a wider range of transparency features for both DBI-level and environment-level, while improving the same time its performance and robustness. While at the moment we use a large set of regression-tests, we believe that we can design a formal approach to prove the soundness of an analysis, or at least a way to profile the DBI itself.

*Concolic Execution and Taint Analysis::* At the moment we are adding support for taint analysis and concolic execution on top of COBAI. We believe that we can manipulate an application whose execution flow relies on certain command line-arguments, or include evasive behaviors. In this way, we may force a desired behavior starting from an initial application execution path (e.g. activate the malicious behavior in sandbox environments, generate a key based on the desired seed, etc.). This is different and more generic compared to current transparency-specific mitigations and may assist the behavior extraction from highly evasive malicious binaries, as well as from APTs (Advanced Persistent Threats) known to target specific conditions to trigger the execution of malicious payloads.

*Automated Tasks::* In order to build a malware removal tool (for persistent, hard to remove malicious applications) or ransomware decryption tool (to decrypt and recover files encrypted by Ransomware), one must analyze malicious samples and then combine the results with the whole process of software development. We plan to reduce and automate the redundant work by interpreting the results of COBAI's analysis and use them to fill templates for such tools. Therefore, an analyst could use COBAI to obtain a malicious trace and also be able to generate a removal tool based on this trace, with minimal effort.

## VI. Conclusion

In this paper, we proposed COBAI as a DBI framework aimed towards mitigating transparency attacks from malicious binaries, while also successfully analyzing their behaviour. Despite common DBI engines which forces developers to combine different tools functionalities in a single one or to rely on the core functionalities of the DBI engine, COBAI has a dynamic architecture supporting multiple tools at once (some of them as plugins), making possible the replacement of core-modules, and at the same time the usage of standalone DBI tools in the same analysis. The dynamic architecture also allows the usage of the DBI as a highly configurable stand-alone tool, able to produce out-of-the box execution traces involving no programming skills. This is different from most existing DBIs like PIN and DynamoRIO, which force users to either use existing open-source DBI-tools to add analysis-functionalities to the DBI-engine, or develop custom ones.

The development of COBAI started in late 2018 as a proof of concept DBI, and since mid 2019 its development and architecture was directed towards the improvement of transparency issues, starting with the DBI itself and extending these features up to the OS environment level in an attempt to automate malicious binary analysis. From our experiments, until now, it was successfully used to reduce ransomware analysis, from days to minutes. Before COBAI was developed, we tested PIN and DynamoRIO, and the negative experiences required to develop and maintain such projects were used to optimize COBAI architecture and usage since the beginning.

Our test-suite both aggregates all of the public available tests (developing DBI-transparency attack scenarios) and also considers new scenarios. While all the transparency issues are covered by problems **P1,P2,P3**, we consider there is still room for unexplored attack scenarios. Uncovering these possibilities in the near future will assist the improvements of automated binary analysis for malicious applications.

Our evaluation for DBI-transparency and environment-transparency issues shows that COBAI is capable of mitigating DBI-transparency issues for $\approx 96\%$ of the tests, and all of the environment-transparency issues described in *pafish* and *al-khaser*. Based on our results, while still aiming toward optimizations and improvements, COBAI might be one of the few fine-grained binary analysis engines, capable of reducing the time involved in reverse-engineering of malicious binaries. The goal is for COBAI to achieve the capability of analyzing complex malicious binaries in any sandbox environment, and maybe new features and performance optimizations will make it more likely to replace conventional analysis tools like debuggers. We believe that technologies developed in COBAI will extend the context of using DBIs for benign and malign binary analysis. Here are several possible scenarios that support our claims:

- The dynamic architecture and highly configurable usage might open the doors for: involving the usage of DBIs in academic research (involve students in extending and optimizing specific core DBI modules); application profiling made easier; system administrators may gather additional logs from applications; easy to add extensions to support a plethora of architectures.
- The transparency module can assist in providing consistent traces for malicious binaries, impossible to achieve with existing technologies.
- The server-client interface can contribute to: network-level debuggers, debugging a countless number of processes and payloads across the network at the same time; the ground for DBI services either offloading part of the DBI analysis to more performant systems, or provide access to real-time execution logs and analysis results.